\documentclass[conference]{IEEEtran}

\usepackage{multicol}
\usepackage{amsmath,amssymb,cite,multirow,subfigure}
\usepackage{graphicx}
\usepackage{url}
\usepackage[linesnumbered,ruled, vlined]{algorithm2e}
\usepackage{algorithmic}
\usepackage{makecell}

\begin{document}
%
\title{A Reduced Latency List Decoding Algorithm for Polar Codes}

\author{\authorblockN{Jun Lin, Chenrong Xiong and Zhiyuan Yan}
\authorblockA{Department of Electrical and Computer Engineering, Lehigh University, PA, USA\\
Email: \{jul311,chx310,yan\}@lehigh.edu}}

\maketitle

\begin{abstract}
The cyclic redundancy check (CRC) aided successive cancelation list (SCL) decoding algorithm has better error performance than the successive cancelation (SC) decoding algorithm for short or moderate polar codes. However, the CRC aided SCL (CA-SCL) decoding algorithm still suffer from long decoding latency. In this paper, a reduced latency list decoding (RLLD) algorithm for polar codes is proposed. For the proposed RLLD algorithm, all rate-0 nodes and part of rate-1 nodes are decoded instantly without traversing the corresponding subtree. A list maximum-likelihood decoding (LMLD) algorithm is proposed to decode the maximum likelihood (ML) nodes and the remaining rate-1 nodes. Moreover, a simplified LMLD (SLMLD) algorithm is also proposed to reduce the computational complexity of the LMLD algorithm. Suppose a partial parallel list decoder architecture with list size $L=4$ is used, for an (8192, 4096) polar code, the proposed RLLD algorithm can reduce the number of decoding clock cycles and decoding latency by 6.97 and 6.77 times, respectively.

\end{abstract}

\section{Introduction}
\label{sec:intro}

Polar codes~\cite{arikan} are a significant breakthrough in coding theory, since it is proved that polar codes can achieve the channel capacity of binary-input symmetric memoryless channels in~\cite{arikan} and any discrete or continuous channel in~\cite{sas_polar}.
Polar codes can be efficiently decoded by the low-complexity successive cancelation (SC) decoding algorithm~\cite{arikan} with complexity of $O(N\log N)$, where $N$ is the block length.

Lots of efforts~\cite{ido_list1, ido_list2} have already been devoted to improve the error-correction performance of polar codes with short or moderate lengths. An successive cancelation list (SCL) decoding algorithm was recently proposed in~\cite{ido_list1}, performs better than the SC decoding algorithm and performs almost the same as a maximum-likelihood (ML) decoder~\cite{ido_list1}. In~\cite{ido_list2}, the cyclic redundancy check (CRC) is used to pick the output codeword from $L$ candidates, where $L$ is the list size. The CRC-aided SCL decoding algorithm performs much better than the SCL decoding algorithm at the expense of negligible loss in code rate. For example, it was shown in~\cite{ido_list2} that the CRC-aided SCL decoding algorithm outperforms the SC decoding algorithm by more than 1 dB when the bit error rate (BER) is on the order of $10^{-5}$ for a polar code of length 2048.

Many research efforts~\cite{low_latency_polar, chuan_polar, ml_ssc, yuan_polar, chuan_low_latency} have been devoted to the reduction of the decoding latency of the SC decoding algorithm.
The simplified successive cancelation (SSC) and the ML-SSC decoding algorithms were proposed in~\cite{low_latency_polar} and~\cite{ml_ssc}, respectively. Both SSC and ML-SSC decoding algorithms can reduce the decoding latency of a SC decoder significantly. However, the reduced latency list decoding algorithm has been rarely discussed in open literature.

In this paper, the algorithms that reduce the latency of list polar decoders are investigated. The main contributions are shown as follows.
\begin{enumerate}
\item A reduced latency list decoding (RLLD) algorithm over LLR domain for polar codes is proposed. The proposed RLLD algorithm deals with rate-0 nodes and part of rate-1 nodes in the same way as the SSC decoding algorithm.
\item A list ML decoding (LMLD) algorithm is proposed to decode the ML and remaining rate-1 nodes. For the list size $L\leq 8$, a hardware friendly simplified LMLD (SLMLD) algorithm is also proposed.
\item For list size $L=4$, an efficient hardware architecture for the proposed SLMLD algorithm is presented. Under a TSMC 90nm technology, at the cost of 1.07 million standard NAND gates, the proposed architecture can achieve a frequency of 400MHz with 4 stage of pipelines.
\item For a partial parallel decoder architecture with $L=4$, it is shown that the RLLD with the SLMLD algorithms can reduce the decoding cycles and latency by 6.97 and 6.77 times, respectively.
\end{enumerate}

\section{Preliminaries}
\label{sec: pre}

\subsection{Polar codes encoding} \label{ssec:polar_encoding}
\begin{figure} [hbt]
\centering
  \includegraphics[width=2.1in]{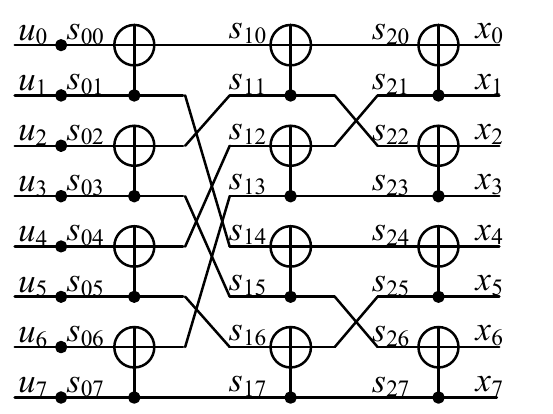}
  \caption{Polar encoder with $N=8$}\label{fig: encoding}
\end{figure}

The generation matrix of a polar code is an $N\times N$ matrix $G=B_NF^{\otimes n}$, where $N=2^n$, $B_N$ is the bit reversal permutation matrix, and $F=\left[{1\atop 1}{0\atop 1}\right]$. Here $\otimes n$ denotes the $n$th Kronecker power and $F^{\otimes n} = F\otimes F^{\otimes (n-1)}$. Let $u_0^{N-1} = (u_0,u_1,\cdots,u_{N-1})$ denote the data bit sequence and $x_0^{N-1} = (x_0,x_1,\cdots,x_{N-1})$ the corresponding encoded bit sequence, then $x_0^{N-1}=u_0^{N-1}G$. The indices of the encoding bit sequence $u_0^{N-1}$ are divided into two sets: the information bits set $\mathcal{A}$ contains $K$ indices and the frozen bits set $\mathcal{A}^c$ contains $N-K$ indices. $u_{\mathcal{A}}$ are the information bits whose indices all come from $\mathcal{A}$. $u_{\mathcal{A}^c}$ are the frozen bits whose indices from $\mathcal{A}^c$. The encoding graph of a polar code with $N=8$ is shown in Fig.~\ref{fig: encoding}.

\subsection{SSC and ML-SSC Decoding Algorithms} \label{ssec: ssc}

\begin{figure} [hbt]
\centering
  \includegraphics[width=2.6in]{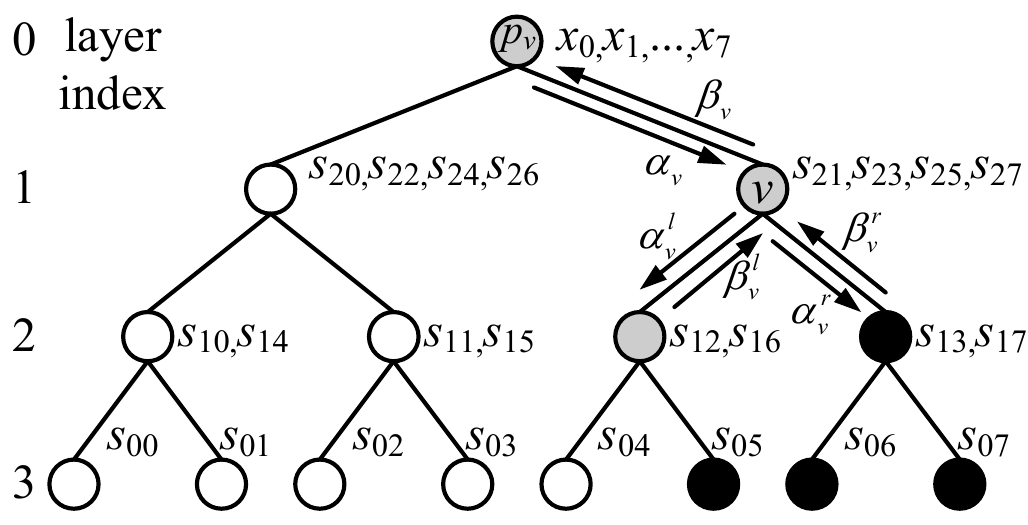}
  \caption{Binary tree representation of a (8, 3) polar code}\label{fig: dec_tree}
\end{figure}

A polar code of length $N=2^n$ can also be represented by a full binary tree of depth $n$~\cite{low_latency_polar}, where each node of the tree is associated with a constituent code. The binary tree representation of an (8, 3) polar code is shown in Fig.~\ref{fig: dec_tree}, where the black and white leaf nodes correspond to information and frozen bits, respectively. In order to show the connection between the tree representation and the direct encoding graph in Fig.~\ref{fig: encoding}, the constituent code associated with each tree node is also shown in Fig.~\ref{fig: dec_tree}. There are three types of nodes in a binary tree representation of a polar code: rate-0 , rate-1 and arbitrary rate nodes. The leaf nodes of a rate-0 and rate-1 nodes are associated with only frozen and information bits, respectively. The leaf nodes of an arbitrary rate node are associated with both information and frozen bits. For example, the rate-0, rate-1 and arbitrary rate nodes in Fig.~\ref{fig: dec_tree} are represented by circles in white, black and gray, respectively.

The SC decoding algorithm can also be mapped on a binary tree, where each node acts as a decoder for its constituent code. As shown in Fig.~\ref{fig: dec_tree}, the decoder at node $v$ receives a soft information vector $\alpha_v$ and returns its correspondent constituent code $\beta_v$. The SC decoding algorithm is initialized by feeding the root node with the channel LLRs, ($L_0, L_1, \cdots, L_{N-1}$), where $L_i = \log(\Pr(y_i|x_i=0)/\Pr(y_i|x_i=1))$. When an internal node $v$ is activated, it calculates the soft information vector $\alpha_{v}^l$ sending to its left child, where
\begin{equation}\label{equ: f_comp}
\alpha_v^l[i] = f(\alpha_v[2i],\alpha_v[2i+1])\mbox{ for } 0\leq i <2^{n-t},
\end{equation}
$f(a,b) = 2\tanh^{-1}(\tanh(a/2)\tanh(b/2))$, and $t$ is the layer index of the child node. $f(a,b)$ can be approximated as:
\begin{equation}\label{equ: f_comp_simplified}
f(a,b)=\mbox{sign}(a)\cdot \mbox{sign}(b)\min(|a|,|b|).
\end{equation}
Node $v$ then waits until it receives the constituent code $\beta_v^l$. The soft information vector
\begin{equation}\label{equ: g_comp}
\alpha_v^r[i] = \alpha_v[2i](1-2\beta_v^l[i])+\alpha_v[2i+1]\mbox{ for } 0\leq i <2^{n-t}.
\end{equation}
Once the right child returns its constituent code $\beta_v^r$, node $v$ computes its constituent code $\beta_v$ as:
\begin{equation}\label{equ: partial_sum}
(\beta_v[2i], \beta_v[2i+1]) = (\beta_v^l[i]\oplus\beta_v^r[i], \beta_v^r[i]),
\end{equation}
where $0\leq i <2^{n-t}$ and $\oplus$ is modulo-2 addition. When a leaf node $v$ is activated, its constituent code $\beta_v$ is set to 0 if leaf node $v$ is associated with a frozen bit. Otherwise, $\beta_v$ is calculated from $\alpha_v$ with the threshold detection:
\begin{eqnarray}\label{equ: hard_dec}
 \beta_v = \left\{ \begin{array}{ll}
 0 & \alpha_v \geq 0\\
 1 & \alpha_v < 0
 \end{array} \right.
\end{eqnarray}
From the root node, all nodes in a tree are activated in a recursive way for the SC decoding. Once $\beta_v$ for the last leaf node is generated, the codeword $x_0^{n-1}$ can be obtained by combining and propagating $\beta_v$ up to the root node.

\begin{figure} [hbt]
\centering
  \includegraphics[width=2.6in]{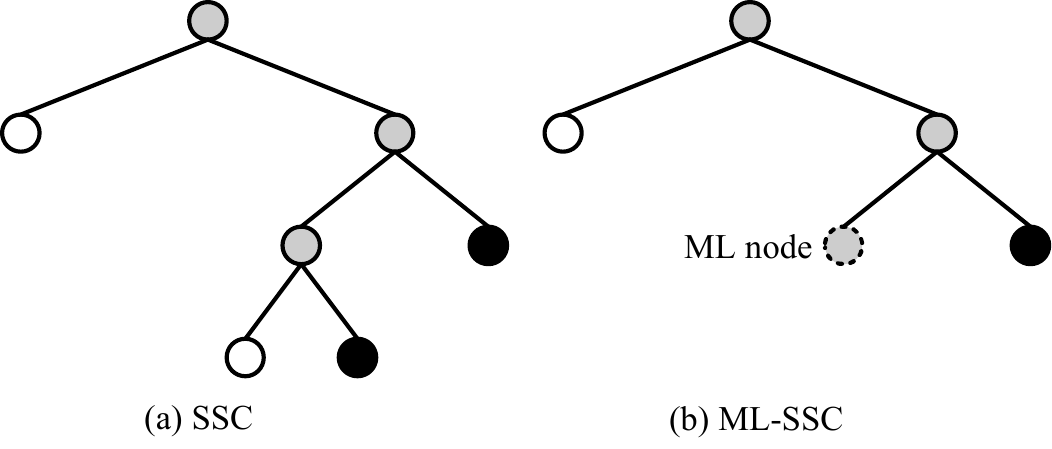}
  \caption{Binary tree representations of a (8, 3) polar code under SSC and ML-SSC decoding algorithms}\label{fig: tree_compare}
\end{figure}

The SSC decoding algorithm in~\cite{low_latency_polar} simplifies the decoding of rate-0 and rate-1 nodes. Once a rate-0 node is activated, it immediately returns its constituent code which is an all zero vector. Once a rate-1 node is activated, its constituent code is directly calculated from the received soft information vector with the threshold detection rule shown in Eq.~(\ref{equ: hard_dec}). The ML-SSC decoding algorithm~\cite{ml_ssc} simplifies the SSC decoding algorithm further by performing the exhaustive-search ML decoding on some resource constrained arbitrary rate nodes, which are called ML nodes in~\cite{ml_ssc}. For an ML node with layer index $t$, the associated constituent code is estimated according to:
\begin{equation}\label{equ: mld}
\beta_v = \arg\!\max_{\textbf{x}\in \mathcal{C}}\sum_{i=0}^{2^{n-t}-1}(1-2\textbf{x}[i])\alpha_v[i],
\end{equation}
where $\mathcal{C}$ is the set of possible constituent codes for the ML node. The binary tree representations of the example (8, 3) polar code under SSC and ML-SSC decoding algorithms are shown in Fig.~\ref{fig: tree_compare}~(a) and (b), respectively. It is observed that the SSC decoding algorithm can reduce the number of nodes to be activated. This number is further reduced by applying the ML-SSC decoding algorithm which introduces ML nodes. It is obvious that all the child nodes of a rate-0 and rate-1 node are still rate-0 and rate-1 nodes, respectively. During the reduction of the binary tree, a rate-0 or rate-1 node is kept only if their parent nodes are not a rate-0 or rate-1 node, respectively. For an arbitrary rate node $v$, let $n_v$ and $d_v$ denote the number of leaf nodes and the number of leaf nodes that correspond to information bits, respectively. In~\cite{ml_ssc}, an arbitrary rate node is labeled as an ML node only if its $n_v$ and $d_v$ do not exceed predefined values.

\subsection{LLR Based List Decoding Algorithms} \label{ssec: llr_based}
In the first several works~\cite{ido_list1, tree_list_dec,jun_list} on list decoding of polar codes, the list decoding algorithm is performed either on probability or logarithmic likelihood (LL) domain. In~\cite{llr_list}, an LLR based list decoding algorithm is proposed to reduce the message memory requirement and the computational complexity of LL based list decoding algorithm. The LLR based list decoding algorithm employs a novel path metric PM$_l^{(i)}$, which is computed as:
\begin{equation}\label{equ: path_metric}
\mbox{PM}_l^{(i)} = \sum_{k=0}^{i}m_i|L_n^{(k)}[l]|,
\end{equation}
where $m_i = 1$ only if $\hat{u}_k[l] = \delta(L_n^{(k)}[l])$ and $\delta(x) = \frac{1}{2}(1-\mbox{sign}(x))$~\cite{llr_list}. Otherwise $m_i = 0$. Here $L_n^{(k)}[l] \triangleq \frac{W_n^{(k)}(\textbf{y},\hat{u}_0^{k-1}[l]|0)}{W_n^{(k)}(\textbf{y},\hat{u}_0^{k-1}[l]|1)}$ and $\textbf{y}$ is the received channel soft information vector.

\section{The Proposed RLLD Algorithm}
\label{sec: rlld}
Though existing list decoding algorithms for polar codes can improve the performance of SC decoders significantly. They still suffer from long decoding latency. During the decoding of each information bit, the current decoding paths need to be doubled and at most $L$ most reliable decoding paths are kept, where $L$ is the list size. The extra cycles spent on path pruning increase the number of the overall decoding cycles~\cite{tree_list_dec}.
In this paper, a reduced latency list decoding (RLLD) algorithm for polar codes is proposed. Let $W_v$ and $I_v$ denote the number of leaf nodes and leaf nodes associated with information bits of a node $v$ in a binary tree, respectively. Let $W_T$ be a predefined threshold value. The general architecture of the proposed RLLD algorithm is shown as follows:
\begin{enumerate}
  \item For a binary tree representation of a polar code, label all the rate-0, rate-1 and ML nodes. For a node $v$ in the tree, let $W_v$ and $I_v$ denote the numbers of leaf nodes and leaf nodes associated with information bits, respectively. For rate-1 nodes, $I_v=W_v$. Moreover, two type of nodes are defined: $T_0$ and $T_1$. $T_0$ nodes include rate-1 nodes with $I_v > W_T$ and all rate-0 nodes. $T_1$ nodes include rate-1 nodes with $I_v \leq W_T$ and all ML nodes. For all ML nodes, $W_v \leq W_{ML}$ and $I_v \leq 8$, where $W_{ML}$ is also a predefined threshold value.
  \item For each decoding path, perform the SC decoding algorithm on the corresponding pruned binary tree, if a $T_0$ node is activated, the corresponding constituent code is decoded immediately and sent to its parent node. Besides, it is unnecessary to compute the LLR vector sent to a rate-0 node, since the constituent code of a rate-0 node is always a zero vector.
  \item If a $T_1$ node is activated, compute $2^{I_v}$ path metrics for each current decoding path, where each path metric corresponds to the reliability of a possible decoding path. Find at most $L$ most reliable decoding paths and continue their corresponding SC decoding. Since only rate-1 nodes with $I_v <W_T$ are involved in the list decoding, the choice of $W_T$ should be decided by the numerical simulation.
  \item Once all $T_0$ and $T_1$ nodes have been activated and all the SC decoding procedures on each decoding path are finished, perform cyclic redundancy check (CRC) on the information bits of each candidate codeword. The output codeword is the one that passes the CRC.
\end{enumerate}

In terms of software or hardware implementation, the proposed RLLD algorithm can be performed over $L$ LLR matrices and $L$ bit matrices. For $l = 0,1,\cdots, L-1$ and $t=1, 2, \cdots, n$, let $P_{l,t}$ be a probability message array of $2^{n-t}$ elements: $P_{l,t}[j]$ stores an LLR message for $j=0,1,\cdots,2^{n-t}-1$. The received channel LLRs are stored in $P_{0,0}$ which has $N=2^n$ elements. $C_{l,t}$ has a similar structure as $P_{l,t}$: $C_{l,t}[j]$ stores two binary partial sums $C_{l,t}[j][0]$ and $C_{l,t}[j][1]$ for $j=0,1,\cdots,2^{n-t}-1$. Let $r_l=(r_l[n-1],r_l[n-2],\cdots,r_l[0])$ be the message updating reference index array for decoding path $l$. For decoding path $l$, $r_l[0]\equiv 0$, while all other elements are initialized with 0. When a $T_0$ or $T_1$ node $v$ is activated, the computation of the soft information vector sent to node $v$ for decoding path $l$ is shown in Algorithm~\ref{algo: soft_vector}, where $t_v$ is the layer index of node $\alpha$ and $P_{l,t_v}$ is the LLR vector sent to node $v$. The $g$ function is shown in Eq.~(\ref{equ: g_comp}). If node $v$ is a rate-0 node, as mentioned before, it is unnecessary to compute the received LLR vector. Under this circumstance,  $t_v$ is decreased by 1. When a decoding path $l$ needs to be copied to decoding path $l'$, the lazy copy approach in~\cite{tree_list_dec} is applied. Instead of copying LLR matrices, $r_l[I_s-1],\cdots, r_l[1]$ are copied to $r_{l'}[I_s-1],\cdots, r_{l'}[1]$, respectively, while $r_{l'}[n],\cdots, r_{l'}[I_s]$ are set to $l'$.

For decoding path $l$, during the computation of $P_{l,t_v}$, LLR arrays, $P_{l,I_s},\cdots, P_{l,t_v}$, need to be updated in serial, where $I_s$ is a pre-computed layer index. For the tree representation of a polar code, suppose all leaf nodes from left to right are indexed from 0 to $N-1$. Let the indices of the leftmost and rightmost leaf nodes of the subtree of node $v$ be IDX$_0$ and IDX$_1$, respectively. $I_s$ is computed based on IDX$_0$ as shown in Algorithm~\ref{algo: I_s_comp}, where the function dec2bin computes the binary representation of its input and $B_{n-1}$ and $B_0$ are the most and least significant bits, respectively.

\begin{algorithm}
\DontPrintSemicolon
\label{algo: soft_vector}
\SetKwInOut{Input}{input}\SetKwInOut{Output}{output}

\Input{$I_s, t_v$}
\Output{$P_{l,t_v}$}
\BlankLine

\For{$t = I_s$ \KwTo $t_v$} {
\For{$k=0$ \KwTo $2^{n-t}$} {
\If{$t == I_s$}{
$b_s=C_{l,t}[k][0]$\;
$P_{l,t}[k] = g(P_{r_l[t-1],t-1}[2k],P_{r_l[t-1],t-1}[2k+1],b_s)$\;
}\Else{
$P_{l,t}[k] = f(P_{r_l[t-1],t-1}[2k],P_{r_l[t-1],t-1}[2k+1])$\;
}
}
}
\caption{llrComp$(l, \alpha)$}
\end{algorithm}

\begin{algorithm}
\DontPrintSemicolon
\label{algo: I_s_comp}
\SetKwInOut{Input}{input}\SetKwInOut{Output}{output}

\Input{IDX$_0$}
\Output{$I_s$}
\BlankLine

\lIf{$\mbox{IDX}_0 == 0$} {
$I_s = 0$
}\Else{
$I_s = n$\;
$(B_n, B_{n-1},\cdots, B_0) = \mbox{dec2bin}(\mbox{IDX}_0)$\;
\For{$j = 0$ \KwTo $n-1$} {
\lIf{$B_j == 0$} {
$I_s = I_s -1$
} \lElse{
\textbf{break}
}
}
}
\caption{}
\end{algorithm}

Once the constituent code $\mathcal{C}_v^l$ sent from node $v$ for decoding path $l$ is computed, $\mathcal{C}_v^l$ is stored in $C_{l,t_v}[k][0]$ for $k=0,1,\cdots, 2^{n-t_v}$ if node $v$ is the left child of its parent node. Otherwise $\mathcal{C}_v^l$ is stored in $C_{l,t_v}[k][1]$ for $k=0,1,\cdots, 2^{n-t_v}$. If the contents of decoding path $l$ need to be copied to decoding path $l'$, the partial sums in decoding path $l$ are copied to the corresponding locations in decoding path $l'$. If node $v$ is the right child of its parent node, then the partial sum computation for path $l$ is performed as shown in Algorithm~\ref{algo: psum_comp}. The input $I_e$ is a layer index and can be obtained by applying Algorithm~\ref{algo: I_s_comp} with IDX$_0$ and $I_s$ being replaced with IDX$_1$ and $I_e$, respectively.

\begin{algorithm}
\DontPrintSemicolon
\label{algo: psum_comp}
\SetKwInOut{Input}{input}\SetKwInOut{Output}{output}

\Input{$I_e, t_v$}
\BlankLine

\For{$t = t_v$ \KwTo $I_e$} {
\For{$k=0$ \KwTo $2^{n-t-1}$} {
\If{$t == I_e$}{
$C_{l,t}[2k][0] = C_{l,t-1}[k][0]\oplus C_{l,t-1}[k][1]$\;
$C_{l,t}[2k+1][0] = C_{l,t-1}[k][1]$\;
}\Else{
$C_{l,t}[2k][1] = C_{l,t-1}[k][0]\oplus C_{l,t-1}[k][1]$\;
$C_{l,t}[2k+1][1] = C_{l,t-1}[k][1]$\;
}
}
}
\caption{pSumComp$(l, \alpha)$}
\end{algorithm}

\subsection{LMLD Algorithms} \label{ssec: mld}
When a $T_1$ node is activated, the current decoding paths will expand, and at most $L$ most reliable decoding paths are kept. In this paper, a list ML decoding (LMLD) algorithm is proposed to find at most $L$ most reliable decoding paths. For a $T_1$ node $v$, there are $2^{I_v}$ candidate output constituent codes since the number of information bits associated with the leaf nodes of a node $v$ is $I_v$. Therefore, for each decoding path $l$, the proposed LMLD algorithm computes $2^{I_v}$ extended path metrics PM$_l^j$ for $j=0,1,\cdots,2^{I_v}-1$ based on the current path metric PM$_l$. Finding the $L$ most reliable surviving decoding paths is equivalent to find the $L$ most reliable constituent codes among all candidates. Here, several conclusions are made on path metrics and extended path metrics:
\begin{itemize}
  \item For each decoding path $l$, the path metric PM$_l$ is initialized with 0. The extended path metrics are computed only when a $T_1$ node is activated.
  \item For each decoding path $l$, each extended path metric PM$_l^j$ corresponds to the reliability measure of the associated candidate constituent code $\mathcal{C}_{v,l}^{j}$ sent from node $v$.
  \item The extended path metric PM$_l^j$ is computed as shown in Eq.~(\ref{equ: epm}), where NM$_l^j$ is called node metric and $\mbox{NM}_l^j = \sum_{k=0}^{2^{n-t_v}-1}m_k|\alpha_{v,l}[k]|$. $\alpha_{v,l}$ is the LLR vector received by the node $v$. $m_k = 1$ only if $\mathcal{C}_{v,l}^{j}[k] = \delta(\alpha_{v,l}[k])$, where $\delta(x)=\frac{1}{2}(1-\mbox{sign}(x))$. Otherwise, $m_k = 0$. As shown in Eq.~(\ref{equ: epm}), for $k=0,1,\cdots,2^{I_v}-1$, if $\mathcal{C}_{v,l}^{j}[k]$ does not equal to the threshold detection based on $\alpha_{v,l}[k]$, then $\mbox{PM}_l^{j}$ is punished by adding the absolute value of $\alpha_{v,l}[k]$. As a result, the smaller a extended path metric is, the more reliable a corresponding constituent code is.
  \begin{equation}\label{equ: epm}
   \mbox{PM}_l^{j} = \mbox{PM}_l + \mbox{NM}_l^j
  \end{equation}
\end{itemize}

Based on the previous conclusions, the proposed LMLD algorithm finds the $L$ most reliable constituent codes by sorting out the $L$ minimum metrics among $2^{I_v}L$ metrics. Let set $\mathbf{S}=\{(l,j)_r|r=0,1,\cdots,L-1\}$, where $(l,j)_r$ is the index of a candidate constituent code. Thus, the proposed LMLD algorithm is shown in Eq.~(\ref{equ: lmld}),
\begin{equation}\label{equ: lmld}
\mathbf{S}={\arg\!\min\!-\!L}_{\substack{l\in[0,L-1]\\j\in[0,2^{I_v}-1]}}\mbox{PM}_l^j,
\end{equation}
where $\arg\!\min\!-\!L$ finds the associated indices of the $L$ minimum metrics among all input metrics. The current $L$ path metrics are updated with the $L$ minimum extended path metrics.

\begin{figure} [hbt]
\centering
  \includegraphics[width=2.6in]{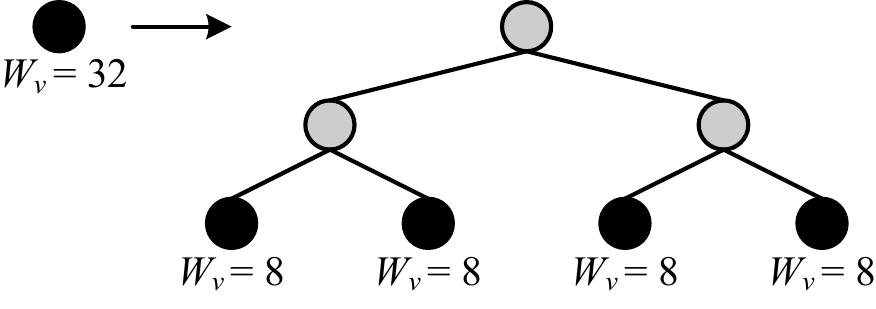}
  \caption{The tree split of a rate-1 node with $W_v=32$}\label{fig: tree_split}
\end{figure}

As shown in Eq.~(\ref{equ: lmld}), the computational complexity of the proposed LMLD algorithm is exponential to $I_v$ which is the number of leaf nodes associated with information bits for node $v$. As a result, the maximum value of $I_v$ should be limited for practical implementation of the proposed LMLD algorithm.
In this paper, the maximum value of $I_v$ is set to 8. The maximum number of leaf nodes of a ML node is set to $W_{ML}=16$. In case of $W_T$ is greater than 8, the corresponding rate-1 node is split to several rate-1 nodes with $W_v= 8$. The other generated nodes due to the split are viewed as arbitrary rate nodes. Take a rate-1 node with $W_v=32$ as an example, the split is shown in Fig.~\ref{fig: tree_split}, where 4 rate-1 nodes with $W_v=8$ are generated while the other generated nodes are deemed as arbitrary rate nodes. Besides, $W_v$ for a rate-1 node can only be a power of 2.

\subsection{SLMLD Algorithms} \label{ssec: smld}
The computational complexity of the proposed LMLD algorithm is still high when $I_v$ is close to 8. In this paper, for $L=\leq 8$, a simplified list ML decoding (SLMLD) algorithm suitable for parallel hardware implementation is proposed to reduce the computational complexity of the proposed LMLD algorithm in further. Here, $L$ is assumed to be a power of 2. The proposed SLMLD algorithm shown in Eq.~(\ref{equ: lmld}) is divided into two major steps:
\begin{enumerate}
  \item For each current decoding path $l$, find its most reliable $L$ constituent codes based on node metrics. Since only the $L$ most reliable constituent codes are needed at last and at most $L$ constituent codes are from the same decoding path $l$, it is enough to find the $L$ most reliable constituent codes for a decoding path $l$.
  \item Compute the extended path metrics based on survived node metrics from previous step, and find the final $L$ most reliable constituent codes based on these $L\times L$ extended metrics.
\end{enumerate}

Depending on the value of $I_v$, the first step can be simplified further. If $2^{I_v}\leq L$, nothing needs to be done. If $2^{I_v} = 2L$, the minimum $L$ extended path metrics and their corresponding $l$ and $j$ indices are computed with a bitonic sequence~\cite{bitonic_sorter} based sorter (BBS)~\cite{jun_list}, where the BBS first transforms the inputs into a bitonic sequence and then generates $L$ minimum metrics among all inputs. When $2^{I_v}>2L$, the minimum $L$ node metrics are computed as follows:
\begin{itemize}
  \item The $2^{I_v}$ node metrics are divided into $L$ groups as follows:
\begin{equation}\nonumber
\underbrace{\mbox{NM}_l^0, \cdots, \mbox{NM}_l^{q-1}}_{\mbox{group 1}},\cdots,\underbrace{\mbox{NM}_l^{(L-1)q}, \cdots, \mbox{NM}_l^{Lq-1}}_{\mbox{group } L},
\end{equation}
  where $q= \frac{2^{I_v}}{L}$. The minimum two metrics of each group are then computed.
  \item Among the resulting $2L$ extended path metrics, the minimum $L$ extended path metrics and their corresponding $l$ and $j$ indices are computed with a BBS.
\end{itemize}
When list size $L=2$, for any $I_v$ values, the first step is just finding the minimum two extended path metrics and their corresponding index pairs $(l,j)$'s.

The second step of the proposed SLMLD algorithm employs the 2$L$-$L$ BBS sorter with $2L$ inputs and $L$ outputs repeatedly to generate $L$ final extended path metrics and their associated path indices. Take $L=4$ as an example, there are $4L$ extended path metrics: PM$_{l_0}^{j_0}$, PM$_{l_1}^{j_1}$, $\cdots$, PM$_{l_{4L-1}}^{j_{4L-1}}$, then PM$_{l_0}^{j_0}$, $\cdots$, PM$_{l_{2L-1}}^{j_{2L-1}}$ and PM$_{l_{2L}}^{j_{2L}}$, $\cdots$, PM$_{l_{4L-1}}^{j_{4L-1}}$ are applied to two 2$L$-$L$ BBSs, respectively. Thus, total $2L$ metrics are selected out. Then the 2$L$-$L$ BBS is employed again to generate the final $L$ minimum extended path metrics: PM$_{l'_0}^{j'_0}$, PM$_{l'_1}^{j'_1}$, $\cdots$, PM$_{l'_{L-1}}^{j'_{L-1}}$.

\subsection{Simulation Results} \label{ssec: sim1}

\begin{figure} [hbt]
\centering
  \includegraphics[width=2.8in]{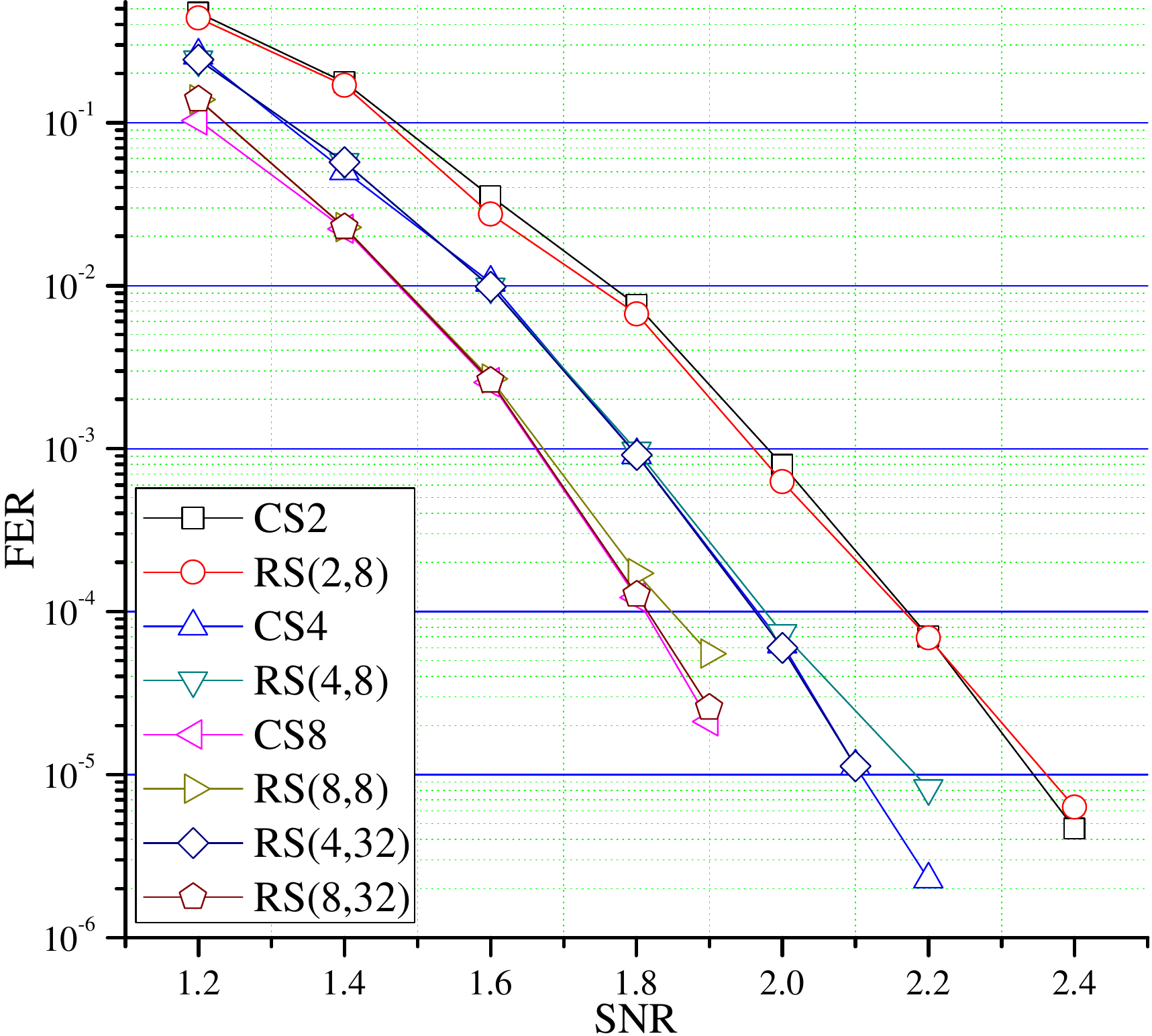}
  \caption{FER performance simulation for an (8192, 4096) polar code}\label{fig: fer8192}
\end{figure}

For an (8192, 4096) polar code, the frame error rate (FER) performance of the proposed RLLD algorithm are shown in Fig.~\ref{fig: fer8192}, under the AWGN channel with BPSK modulation. As shown in Fig.~\ref{fig: fer8192}, CS$i$ denotes the CRC aided SC list decoding algorithm~\cite{ido_list1} with list size $L=i$ over LLR domain, and RS($i,\omega$) denotes the proposed RLLD algorithm with the SLMLD algorithm when list size $L=i$ and $W_T = \omega$. For both CS$i$ and RS($i, \omega$) algorithms, 32 information bits are replaced with a 32-bit CRC checksum.

For simplicity, the FER performances of the proposed RLLD algorithm with LMLD (RL) algorithm are not shown in this paper, since the FER performances of the RL algorithm are the same as that of the CS algorithm with the same list size.

Based on the simulation results, the following conclusions are made:
\begin{itemize}
  \item The performance of the proposed RS algorithm is affected by the list size $L$. For the (8192, 4096) polar code, the FER performances of RS(2, 8) is close that of CS2. However, RS(4, 8) and RS(8, 8) show performance degradation when the FER is blow $10^{-4}$.
  \item In order to achieve good error correction performance, for the proposed RS algorithm, the threshold value $W_T$ should be large enough. A larger $W_T$ will transfer more rate-1 nodes to $T_1$ nodes, which in turn increases the chance that a correct codeword shows in the final lists. For the (8192, 4096) polar codes, RS(4, 8) and RS(8, 8) perform worse than RS(4, 32) and RS(8, 32), respectively, when the SNR is large.
  \item The side effect of increasing $W_T$ is that both the decoding complexity and latency will increase since more $T_1$ nodes are generated. Based on simulation results shown in Fig.~\ref{fig: fer8192}, a dynamic $W_T$ can be adopt for the proposed RS algorithm in order to achieve the most latency reduction at different SNR regions while maintaining the error correction performance.
\end{itemize}

\subsection{Hardware implementation of the proposed SLMLD} \label{ssec: slmld_archi}
\begin{figure} [hbt]
\centering
  \includegraphics[width=2.8in]{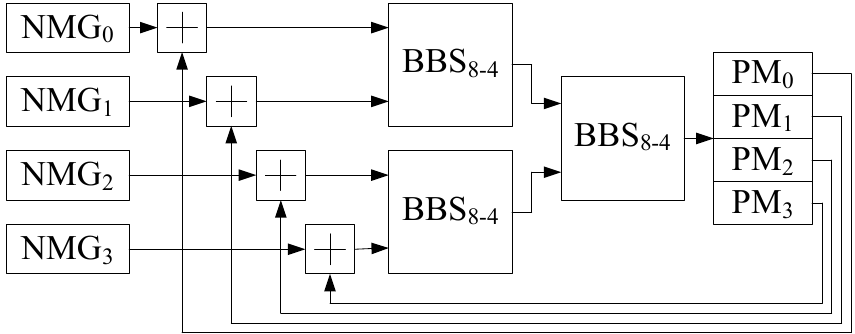}
  \caption{The proposed architecture for SLMLD}\label{fig: slmld}
\end{figure}

In this paper, an efficient hardware implementation of the proposed SLMLD algorithm is shown in Fig.~\ref{fig: slmld}, where the corresponding list size $L=4$, and the architectures for other $L$ values can be inferred. As shown in Fig.~\ref{fig: slmld}, the node metric generation (NMG) unit finds $L$ minimum node metrics and their corresponding constitution codes for each decoding path. For the decoding path $l$, the extended path metrics PM$_l^j$'s are obtained by adding the node metrics with the path metric PM$_l$, which is stored in registers and initialized with 0. BBS$_{8-4}$ in Fig.~\ref{fig: slmld} denotes the BBS with 8 metrics to be sorted. Two stages of BBS$_{8-4}$ find the 4 minimum extended path metrics and their corresponding constituent codes and list indices.

\begin{figure} [hbt]
\centering
  \includegraphics[width=2.8in]{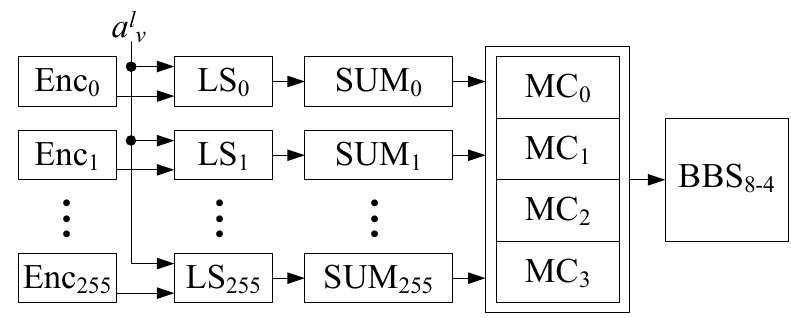}
  \caption{Hardware architecture of the proposed NMG unit}\label{fig: nmg}
\end{figure}

The hardware architecture of the NMG unit is shown in Fig.~\ref{fig: nmg}. Since the maximum value of $I_v$ is 8 for any $T_1$ node, there are at most $2^8=256$ candidate constituent codes for a $T_1$ node $v$. Each Enc unit in Fig.~\ref{fig: nmg} is responsible for generating a candidate constituent code based on the encoding of polar codes. For $j=0,1,\cdots,2^{I_v}-1$, the LLR selection unit, LS$_j$, and the summation unit, SUM$_j$, work together to compute the node metric NM$_l^j$ shown in Eq.~(\ref{equ: epm}). Based on the input LLR vector $\alpha_{v,l}$, LS$_j$ outputs an LLR vector which has the same amount of elements as that of $\alpha_{v,l}$. For $k=0,1,\cdots,2^{n-t_v}-1$, the $k$-th output LLR is 0 only if $m_k =0 $. Otherwise, the output LLR is $|\alpha_{v,l}[k]|$. The SUM$_j$ unit just adds up all its input LLRs sent from LS$_j$ and outputs the corresponding node metric. The minimum two LLRs computation (MC) unit in Fig.~\ref{fig: nmg} finds out the first and the second minimum LLRs and their corresponding constituent codes among all its inputs. When $L=4$, as shown in Fig.~\ref{fig: nmg}, the computed node metrics are divided into 4 groups and fed to 4 MC units, respectively. The BBS$_{8-4}$ unit generates 4 finally survived node metrics and their corresponding constituent codes.

In this paper, the proposed architecture for the SLMLD algorithm is synthesized under a TSMC 90nm CMOS technology. With 4 stages of pipeline registers, it achieves a frequency of 400MHz and consumes 1.07 million standard NAND gates. For our implementation, when a $T_1$ node is activated, it will take 4 clock cycles to find the surviving constituent codes and decoding paths. The area of the architecture of the SLMLD algorithm is almost the same as an LLR based list decoder with $L=4$.

\subsection{Comparisons of decoding clock cycles and latency} \label{ssec: cycle_cmp}

Since the detailed decoding cycles of list decoders are related with a detailed hardware architecture, in this paper, the decoding latency comparison is performed based on the assumption that the partial parallel list architecture~\cite{tree_list_dec} is employed and there are $P=128$ processing units for each decoding path. Let $N_R$ denote the clock cycles used to decode a codeword for decoders with the proposed RS algorithm. Then $N_R = N_L+N_P$, where $N_L$ and $N_P$ are cycles used on the LLR computation and path pruning, respectively. Besides, $N_P=N_aN_s$, where $N_a$ is the times that a $T_1$ node is activated and $N_s$ is the number pipelines inserted in the implementation of the SLMLD algorithm. Let $N_C$ denote the clock cycles used to decode a codeword for decoders with the CS algorithm. Then $N_C=2N+\frac{N}{P}\log_2(\frac{N}{4P})+NR$~\cite{tree_list_dec}, where $N$ and $R$ are the code block length and rate, respectively.

For the aforementioned (8192, 4096) polar code used in our simulations in Section~\ref{ssec: sim1}, $N_L=1207$, $N_a=441$ and $N_s=4$ when $W_T=32$ and $L=4$. Thus, $N_R=1207+441\times 4=2971$. Meanwhile, the cycles $N_C=2\times 8192+\frac{8192}{128}\times \log_2(\frac{8192}{512})+4096=20736$. With the proposed RS decoding algorithm, the clock cycles used for decoding one codeword is reduced by about 6.97 times.

Under the UMC 90nm CMOS technology, the (8192, 4096) list polar decoder can achieve a frequency of 412MHz~\cite{llr_list} when list size $L=4$. Since the list decoder with the proposed RS decoding algorithm need only to change the path pruning part, the proposed list decoder can only achieve a frequency of 400MHz under 90nm technology. Thus, the decoding latency is reduced by about 6.77 times due to the proposed RS decoding algorithm when $L=4$.

\section{Conclusion}
In this paper, a reduced latency decoding algorithm for polar codes is proposed. The hardware implementation of the SLMLD is also discussed. The future work includes studying the performances of the proposed RLLD algorithm when FER is below $10^{-10}$. Besides, more efficient implementations of the proposed SLMLD algorithm when list size is large will be investigated.
\bibliographystyle{IEEEbib}
\bibliography{refs}

\begin{thebibliography}{10}

\bibitem{arikan}
E.~Ar{\i}kan,
\newblock ``Channel polariztion: a method for constructing capacity-achieving
  codes for symmetric binary-input memoryless channels,''
\newblock {\em IEEE Trans. Info. Theory}, vol. 55, no. 7, pp. 3051--3073, Jul.
  2009.

\bibitem{sas_polar}
{E. Sasoglu, E. Teltar and E. Ar{\i}kan},
\newblock ``Polariztion for arbitrary discrete memoryless channels,''
\newblock in {\em Proc. IEEE Int. Symp. on Information Theory}, 2009, pp.
  144--148.

\bibitem{ido_list1}
I.~Tal and A.~Vardy,
\newblock ``List decoding of polar codes,''
\newblock in {\em Proc. IEEE Int. Symp. on Information Theory}, Jul. 2011, pp.
  1--5.

\bibitem{ido_list2}
I.~Tal and A.~Vardy,
\newblock ``List decoding of polar codes,''
\newblock in {\em \url{http://arxiv.org/abs/1206.0050}}.

\bibitem{low_latency_polar}
A.~Alamdar-Yazdi and F.~R. Kschischang,
\newblock ``A simplified successive-cancellation decoder for polar codes,''
\newblock {\em IEEE Commun. Lett.}, vol. 15, no. 12, pp. 1378--1380, Dec. 2011.

\bibitem{chuan_polar}
C.~Zhang and K.~K. Parhi,
\newblock ``Low-latency sequential and overlapped architectures for successive
  cancellation polar decoder,''
\newblock {\em IEEE Trans. Signal Processing}, vol. 61, no. 10, pp. 2429--2441,
  Mar. 2013.

\bibitem{ml_ssc}
G.~Sarkis and W.~J. Gross,
\newblock ``{Increasing the Throughput of Polar Decoders},''
\newblock {\em IEEE Commun. Lett.}, vol. 17, no. 9, pp. 725--728, Apr 2013.

\bibitem{yuan_polar}
B.~Yuan and K.~K. Parhi,
\newblock ``Low-latency successive-cancellation polar decoder architectures
  using 2-bit decoding,''
\newblock {\em {IEEE} Trans. on Circuits Syst. I, Reg. Papers}, {to appear}.

\bibitem{chuan_low_latency}
C.~Zhang and K.~K. Parhi,
\newblock ``Latency analysis and architecture design of simplified sc polar
  decoders,''
\newblock {\em {IEEE} Trans. on Circuits Syst. II, Exp. Briefs}, vol. 61, no.
  2, pp. 115--119, Feb. 2014.

\bibitem{tree_list_dec}
{A. Balatsoukas-Stimming, A. J. Raymond, W. J. Gross and A. Burg},
\newblock ``{Tree search architecture for list SC decoding of polar codes},''
\newblock in {\em \url{http://arxiv.org/abs/1303.7127}}.

\bibitem{jun_list}
J.~Lin and Z.~Yan,
\newblock ``Efficient list decoder architecture for polar codes,''
\newblock in {\em Proc. IEEE Int. Symp. on Circuits and Systems (ISCAS)}, Jun.
  2014, to appear.

\bibitem{llr_list}
{A. Balatsoukas-Stimming, M. B. Parizi and A. Burg},
\newblock ``{LLR-Based Successive Cancellation List Decoding of Polar Codes},''
\newblock in {\em \url{http://arxiv.org/pdf/1401.3753v1.pdf}}.

\bibitem{bitonic_sorter}
K.~E. Batcher,
\newblock ``Sorting networks and their applications,''
\newblock in {\em Proc. ACM spring joint computer conference}, Apr. 1968, pp.
  307--314.

\end{thebibliography}

\end{document}